# Chip-based photon quantum state sources using nonlinear optics

Lucia Caspani,[1,2] Chunle Xiong,[3] Benjamin J. Eggleton,[3] Daniele Bajoni,[4] Marco Liscidini,[5] Matteo Galli,[5] Roberto Morandotti,[6,7,8] and David J. Moss[9]

[1]Institute of Photonics, Department of Physics, University of Strathclyde, Glasgow G1 1RD, UK.

[2]Institute of Photonics and Quantum Sciences, Heriot-Watt University, Edinburgh EH14 4AS, UK.

[3]Centre for Ultrahigh bandwidth Devices for Optical Systems (CUDOS), Institute of Photonics and Optical Science (IPOS), School of Physics, University of Sydney, NSW 2006, Australia.

[4]Dipartimento di Ingegneria Industriale e dell'Informazione, Università di Pavia, via Ferrata 1, 27100, Pavia, Italy.

[5]Dipartimento di Fisica, Università di Pavia, via Bassi 6, 27100 Pavia, Italy.

[6]INRS-EMT, 1650 Boulevard Lionel-Boulet, Varennes, Québec J3X 1S2, Canada.

[7]Institute of Fundamental and Frontier Sciences, University of Electronic Science and Technology of China, Chengdu 610054

[8]National Research University of Information Technologies, Mechanics and Optics, St. Petersburg, Russia.

[9]Center for Microphotonics, Swinburne University of Technology, Hawthorn, Victoria, 3122 Australia.

**Corresponding Author:** Lucia Caspani, Institute of Photonics, Department of Physics, University of Strathclyde, Glasgow G1 1RD, UK, Email: lucia.caspani@strath.ac.uk
**Corresponding Author:** David Moss, Center for Microphotonics, Swinburne University of Technology, Hawthorn, Victoria, 3122 Australia, Email: dmoss@swin.edu.au

**Abstract**
The ability to generate complex optical photon states involving entanglement between multiple optical modes is not only critical to advancing our understanding of quantum mechanics but will play a key role in generating many applications in quantum technologies. These include quantum communications, computation, imaging, microscopy and many other novel technologies that are constantly being proposed. However, approaches to generating parallel multiple, customisable bi- and multi-entangled quantum bits (qubits) on a chip are still in the early stages of development. Here, we review recent developments in the realisation of integrated sources of photonic quantum states, focusing on approaches based on nonlinear optics that are compatible with contemporary optical fibre telecommunications and quantum memory infrastructures as well as with chip-scale semiconductor technology. These new and exciting platforms hold the promise of compact, low-cost, scalable and practical implementations of sources for the generation and manipulation of complex quantum optical states on a chip, which will play a major role in bringing quantum technologies out of the laboratory and into the real world.

**1) Introduction**
Quantum mechanics underpins many of the scientific and technological advancements that have already had a significant impact on our society, ranging from ultrafast computing to high-sensitivity metrology and secure communications. Furthermore, it holds the promise of profound future innovations that will redefine many areas, such as quantum computing, offering unprecedented computational power, as well as emerging areas such as non-classical imaging and spectroscopy, where quantum mechanics offers a means to greatly increase sensitivity. In particular, the field of quantum telecommunications is already providing ultimate communications security that is directly guaranteed by the laws of physics rather than by complex mathematical algorithms.

Most of these technologies exploit the peculiar properties of quantum mechanics, such as the principles of superposition and entanglement. Superposition allows a quantum system to be in two different states simultaneously, while a quantum system composed of more than one component (e.g., particles or photons) is said to be entangled if it can only be described as a whole (see Supplementary Information, Sec. A).

While many different physical systems have been exploited for quantum technologies, including trapped ions and semiconductor circuits, photonics has played a particularly crucial role[1–3]. Historically, light and its ultimate constituent – the photon, or the quantum of light – have served as a testing ground for many breakthrough experiments aimed at confirming the apparent counterintuitive nature of quantum mechanics. This was highlighted by the seminal work on the violation[4] (and more recently, loophole-free violation[5,6]) of Bell's inequalities, which demonstrated the non-local character of quantum mechanics, a fundamental property that cannot be explained by hidden-variables theories, as was suggested 40 years earlier by Einstein, Podolsky and Rosen[7].

Photonics has become a widespread platform for quantum experiments for several reasons: i) the possibility of easily transmitting quantum states encoded in a photon by means of free space optical links or through fibre optic networks; ii) the advances in nonlinear optics that have enabled the generation of single and entangled photons; and iii) the lack of extreme sensitivity to environmental noise (thermal, electromagnetic, etc.) that plagues solid-state approaches. Nonlinear parametric processes have been instrumental in generating fundamental quantum states of light when an intense pump laser field propagates through a nonlinear medium, there is a probability that two new photons are generated as a pair, either as uncorrelated photons or in an entangled state.

The ability to achieve these functions on photonic integrated chips or circuits is absolutely key to moving quantum technologies out of the laboratory and into the real world. The main components of quantum photonic systems, such as mirrors, beam splitters, and phase shifters, are all now realisable in integrated form[8,9]. Ultimately, all functions needed for quantum demonstrations – the generation, manipulation and detection of single/entangled photons – would ideally be integrated in just one chip[10]. However, even just the ability to integrate one function, such as the source of non-classical light, would already offer many advantages over bulk optical setups.

Here, we review recent advances in integrated, or chip-based, sources of quantum states of light, including single and entangled photons, and the techniques for characterising heralded and entangled photon sources. We focus on devices based on nonlinear optics that are compatible with electronic on-chip technology (Complementary Metal Oxide Semiconductor (CMOS)), ending with a discussion on recent achievements in the generation of single photons on demand. We refer the reader elsewhere for other relevant results based on integrated chips, e.g., quantum states[11–14], quantum interference[15–21], quantum logic ports[12,22,23], quantum algorithms[24], quantum walks[25–29], and boson sampling[30–34], and reviews on related topics, including quantum metrology[35], computing[36], integrated detectors, superconducting nanowires [37,38], sources based on a range of different platforms (e.g., GaAs[39], silicon-on-insulator[40], diamond[41] and silicon nitride[42]) and others [9,14,43–51].

**2) Entangled and single-photon sources**
The key states of interest for quantum photonic devices are single and entangled photons. These can be both produced via spontaneous nonlinear parametric processes. Depending on the platform material, these occur via second- ($\chi^{(2)}$) or third-order ($\chi^{(3)}$) nonlinearities, where either one (for $\chi^{(2)}$) or two (for $\chi^{(3)}$) photons from an intense pump laser are annihilated into two daughter photons. The $\chi^{(2)}$ process is termed spontaneous parametric down-conversion (SPDC), while the $\chi^{(3)}$ process is called spontaneous four-wave mixing (SFWM). These processes are the quantum counterparts to the classical difference-frequency generation and four-wave mixing (FWM), respectively. In the non-classical case, the seed fields are provided by vacuum fluctuations: only the virtual signal and idler pairs that satisfy energy and momentum conservation are efficiently transformed into real photons. Alternatively, we can think of SPDC as a photon fission process, while SFWM is more of an elastic scattering process.

One of the main differences between SPDC and SFWM is that for SPDC, energy conservation requires the signal and idler daughter photons to be generated at frequencies that are symmetrically located with respect to half of the pump field frequency, while in SFWM, they are symmetrically distributed around the pump frequency:

$$SPDC: \begin{cases} \omega_s = \omega_p/2 + \Delta\Omega \\ \omega_i = \omega_p/2 - \Delta\Omega \end{cases}; \quad SFWM: \begin{cases} \omega_s = \omega_p + \Delta\Omega \\ \omega_i = \omega_p - \Delta\Omega \end{cases}; \quad (1)$$

where $\omega_p$, $\omega_s$, and $\omega_i$ represent the pump, signal, and idler frequencies, respectively, while $\Delta\Omega$ is the frequency shift with respect to the degenerate process. This implies that in SFWM, all of the involved fields can have similar wavelengths. While this can be useful in satisfying phase matching conditions (momentum conservation), it also increases the difficulty in filtering out the pump to isolate the signal and idler photons.

*2A. Entangled photons:* The combination of vacuum fluctuations and conservation laws is at the core of the entanglement between signal and idler photons. Depending on the configuration of the conversion process, entanglement can be generated in different degrees of freedom, e.g., polarisation, space, time, and orbital angular momentum, and is a fundamental resource for quantum computing and communications. Indeed, many quantum algorithms rely on entanglement[52].

To achieve entanglement, the signal and idler photons need to be generated in at least a two-mode state, e.g., with horizontal and vertical polarisations. For type I SPDC, the signal and idler photons are always generated with the same polarisation, e.g.:

$$|\psi\rangle = |H\rangle_s|H\rangle_i, \quad (2)$$

whereas for type II SPDC, they are generated with orthogonal polarisations, and it is thus possible to obtain, for example, the entangled state:

$$|\psi\rangle_{ent} = |H\rangle_s|V\rangle_i + |V\rangle_s|H\rangle_i . \quad (3)$$

More formally, the two cases are referred to as one- and two-mode squeezing transformations.

Protocols based on entanglement have been proposed (e.g., the E91 protocol[53]) for applications in quantum cryptography, where "Alice" and "Bob" each share a component of a bipartite entangled state. Eavesdropping can be detected by exploiting the collapse of the wave function upon measurement. The multimode nature of the relevant variable provides the alphabet for the exchange of a cryptographic key. The higher the dimensionality of the state, the larger the amount of information each qubit can contain. Different degrees of freedom have been investigated for this purpose, e.g., space[54], time[3,55] (or its conjugate variable, frequency[56]) and orbital angular momentum[57].

*2B. Heralded single photons:* A single photon is a particular quantum state where one and only one photon is present, and it is fundamental for quantum information and computing. One of the most widespread quantum cryptographic protocols, the BB84[58], relies on single photons, where security is provided by the fact that i) it is not possible to measure the quantum state of a system without perturbing it; ii) a single photon cannot be partially measured since it is the ultimate quantum of electromagnetic radiation; and iii) it is not possible to perfectly clone an unknown quantum state (no-cloning theorem[59,60]). In 2000, a universal quantum computing approach based on single photons and linear optics[61] was proposed, commonly referred to as linear optical quantum computing (LOQC). For all these applications, there is a great need for more efficient and reliable single-photon sources.

Single-photon sources can be distinguished according to whether they are deterministic or probabilistic sources, depending on whether the photons are available "on demand" or at an unknown time, respectively. For cryptography or computing, deterministic sources are much more preferable; these are discussed in Section 5.

In both SPDC and SFWM, the signal and idler photons are always emitted in pairs and correlated in time. This correlated emission, while probabilistic, can be exploited in a heralding scheme where one photon signals the presence of the other, although this approach is limited by both loss and multiple pair generation. Each time a signal or idler photon is lost, either no heralding occurs, and thus the single photon is present but not usable, or vice versa – an empty state is heralded. The state generated by spontaneous parametric processes can in general be expressed as[62]:

$$|\psi\rangle_{SPDC/SFWM} = \sum_{n=0}^{\infty} c_n |n\rangle_s |n\rangle_i, \qquad (4)$$

where $n$ is an integer number, $s$ and $i$ represent signal and idler, respectively, and $c_n = \frac{(\tanh r)^n}{\cosh r}$ represents complex coefficients, with $r$ being a squeeze parameter that depends on the pump intensity (and determines the average photon number $\langle n \rangle$). The probability to find exactly $n$ photons in the signal and $n$ photons in the idler is given by $|c_n|^2$. For vacuum squeezed states, the photon number distribution ($P_n = |c_n|^2 = |(\tanh r)^n / \cosh r|^2$) is maximum at $n=0$, while for other states, such as coherent states, the photon number distribution peaks at $\langle n \rangle$. If the parameter $r$ is small enough (i.e., if the pump intensity is sufficiently low), only the first two terms are relevant, corresponding to either no generation or the generation of a single pair. If multiple pairs are created, more than one photon is simultaneously present in each beam, which can result in the heralding of more than one photon, in turn compromising, for example, quantum cryptography security. As a rule of thumb, the pump intensity should be kept low enough to have an average of no more than 0.1 signal/idler pairs per pump pulse (or per pump coherence time in the case of continuous wave excitation). While this low-gain regime is necessary for heralded single-photon sources, quantum entanglement between signal and idler fields can also be preserved in the high-gain regime, where very intense beams can be generated, as in the case of intensity/phase entanglement in twin beams[63]. By judicious engineering of a probabilistic source, for example, by properly combining different SPDC or SFWM processes, an almost deterministic single-photon source can be realised (see Section 5).

## 3) Characterising a heralded single-photon source
*3A. True single photons:* The key issue with heralded single-photon sources is whether or not the heralded state is indeed a single photon. This is typically determined by measuring the degree of second-order coherence, or the $g^{(2)}(\tau)$ function[62,64], that characterises the photon statistics of a field and that is related to its temporal intensity fluctuations via:

$$g^{(2)}(\tau) = \frac{\langle I(t)I(t+\tau)\rangle}{I^2}, \quad (5)$$

where *I(t)* is the field intensity at time *t* (defined as the average over many field oscillations). It can be measured, for example, by splitting a beam using a 50/50 splitter and then recording the intensity correlations at the output ports as a function of the relative delay (Hanbury-Brown and Twiss, or intensity interferometer).

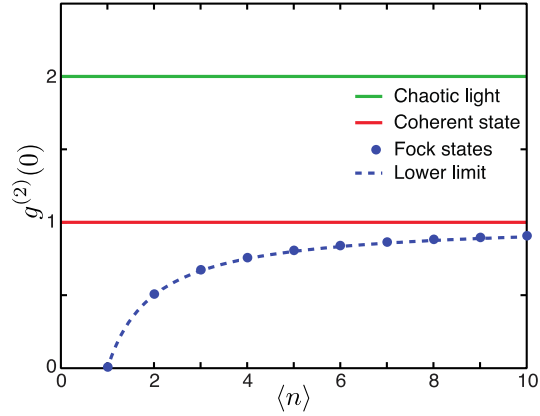

**Figure 1.** Value of $g^{(2)}(0)$ for different states as a function of the average photon number $\langle n \rangle$: chaotic or thermal light (green line), coherent state (red line), and Fock states (blue dots). The dashed blue line represents the lower limit for $g^{(2)}(0)$ in the quantum treatment.

Classically, the value at zero delay is $\geq 1$, i.e., $g^{(2)}_{class}(0) \geq 1$. However, in the quantum treatment, the operator character of the fields must be taken into account; this allows one to access an additional range of values below unity. For example, for Fock, or number, states composed of an exact number of photons (without any intensity fluctuations), we have:

$$g^{(2)}(0) = 1 - \frac{1}{n}, \quad (6)$$

where *n* is the number of photons. A plot of $g^{(2)}(0)$ for different states is shown in Fig. 1.

For a perfect single-photon source, $g^{(2)}(0) = 0$, which can be intuitively understood by considering a single photon entering a 50/50 beam splitter (Fig. 2a). Since a single photon is the ultimate quantum of radiation, it cannot be split further; thus, it can only exit one port of the beam splitter, not both. Therefore, the number of coincidences at the output ports of a beam splitter, as a function of the relative arrival time of photons, displays a dip at zero delay (Fig. 2b). At large delays, $g^{(2)}(\tau)$ approaches unity, regardless of the photon state. The closer the dip is to zero at zero delay, the better the source approaches a true single-photon source. In general, for realistic sources, $g^{(2)}(0) < 0.5$ is required to claim a single-photon state since the theoretical value of $g^{(2)}(0)$ for a two-photon Fock state is 0.5. For a heralded single-photon source, the characterisation setup is very similar, but the coincidences at the beam splitter output are only measured when the heralded photon is detected (Fig. 2c).

*3B. Purity of the state:* In general, a fundamental requirement for a single-photon source is the purity of the generated state. Indeed, many quantum information applications (e.g., LOQC gates[65]) are based on the interference of two or more single photons and require pure states for optimal visibility. Thus, unentangled photons are generally required since this is a necessary condition to herald single photons in a pure state[66]. This situation is in contrast to the generation of entangled photons (see Section 2A), in which quantum correlations are desired and are, in fact, a fundamental property.

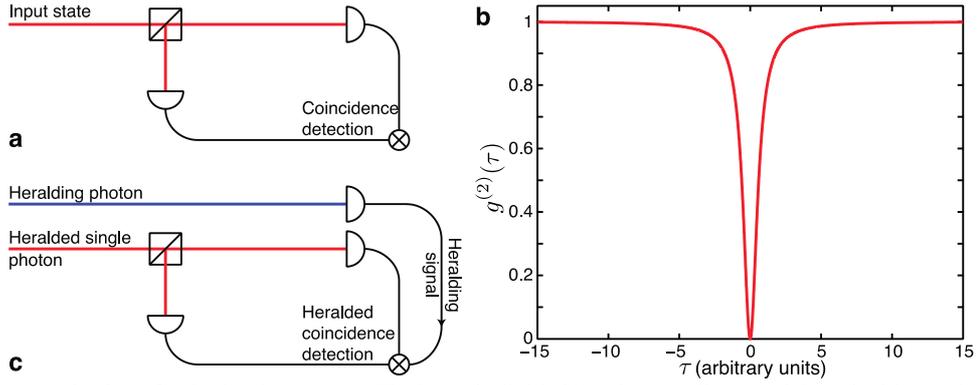

**Figure 2.** (a) Characterisation of a single-photon state. The beam is divided by a beam splitter, and the coincidences between the output ports are recorded as a function of the relative delay τ. (b) Expected second-order coherence function for a single-photon state. At zero delay, we have a dip reaching zero. Note that the shape and width of the function are arbitrary and in general depend on the particular process considered for generating the single photons. (c) Characterisation of a heralded single-photon source. In this case, the coincidences between the output ports of the beam splitter are measured if and only if the detector on the heralding arm fires.

The purity of a single-photon state can be measured using different techniques. The most formal techniques rely on measuring the density matrix of the state, $\hat{\rho}$, using the purity obtained from the trace of the density matrix squared: $\gamma = Tr(\hat{\rho}^2)$, where $\gamma = 1$ refers to a pure state. Generally, this is the most complete characterisation of a quantum state, as it contains all the relevant information for both single photons and entangled states[67,68]. However, determining $\hat{\rho}$ requires several different measurements. For example, for a $D$-dimensional, $n$-partite (e.g., composed of $n$ photons) quantum system, $\hat{\rho}$ is represented by a $D^n \times D^n$ complex matrix. Considering that the density matrix is normalised and Hermitian, i.e., the conditions $Tr(\hat{\rho}) = 1$ and $\hat{\rho} = \hat{\rho}^\dagger$ must hold, it is implied that, in general, $D^{2n}-1$ parameters must be identified. These parameters can be obtained by taking a set of $D^{2n}$ different projection measurements[69]. For example, the state of 2 polarisation-entangled qubits can be characterised by measuring the coincidences in 16 different combinations of the two photon polarisation states (e.g., all combinations of the horizontal, vertical, +45°, and right circular polarisation settings)[69]. Similarly, 3-photon polarisation-entangled states require one to measure triple coincidence events in 64 different settings, and so on.

An alternative approach relies on demonstrating that the source is single mode, since in this case the measurement of the heralding photon will project the single photon into the corresponding pure single mode[70] (see Supplementary Information, Sec. B). Note also that the normalisation condition $Tr(\hat{\rho}) = 1$ combined with the purity condition $Tr(\hat{\rho}^2) = 1$ implies that for a pure state, the diagonalization of the density matrix leads to only 1 non-zero eigenvalue, i.e., a pure state can always be represented by a single-mode state in the proper basis. A single-mode photon can be obtained via a multimode generation process, provided that suitable filtering is applied before detection, although at the expense of reducing the efficiency of the source. Alternatively, single-mode emission can be obtained by modifying the process parameters, such as the pump spectrum and phase matching curve (see Chapter 11.2.4. in[71] for details on heralding pure single-photon states).

The number of modes can be obtained directly by measuring the signal-idler correlations for a specific variable. For example, the single- or multimode character in the frequency domain can be determined by measuring the signal/idler joint spectral distribution (JSD), i.e., the frequency of the idler given the frequency of the signal. Single-mode emission will then be characterised by uncorrelated signal and idler photons (Fig. 3a), while correlation is an indication of a multimode character (Fig. 3b). The JSD can be obtained by measuring, for each idler frequency, the coincidences for all the signal frequencies. This measurement is typically obtained by exploiting narrowband filters (able to resolve the frequency bandwidth over which the signal and idler photons are generated), although this typically introduces significant loss, particularly for very narrow bandwidths. In turn, this can jeopardise the whole measurement by requiring extremely long integration times to compensate for losses. A possible solution is to exploit the corresponding SPDC and SFWM stimulated processes[72,73], for example, by providing as the input the signal field at different frequencies and measuring the idler power. The stimulated process avoids the need for single-photon detectors and strongly reduces the measurement time. This is particularly suitable for characterising states generated by integrated resonators, where the very narrow linewidth requires resolutions of picometres or less and low loss filters are generally not available. Finally, by exploiting the known statistics of the separate signal and idler beams, one can avoid the need for filtering the signal and idler fields, which is particularly useful for very narrow linewidths. In SPDC and SFWM signal and idler beams individually exhibit thermal statistics as a result of the amplification of vacuum fluctuations. In turn, the number of modes of a thermal state can be measured based on the degree of second-order coherence, the zero-delay value of which is related to the number of modes through the relation[64,74,75]:

$$g^{(2)}_{thermal}(0) = 1 + \frac{1}{M}, \quad (7)$$

where $M$ represents the total number of modes of all involved variables. Provided that all the modes are effectively coupled to the detector, note that this technique can resolve very narrow frequency modes. Indeed this requires the temporal resolution of the detector (typically limited by jitter and being of the order of hundreds of picoseconds for telecom detectors) to be shorter than the photon coherence time (which, in turn, is quite long for narrow frequency bandwidth photons, e.g., nanoseconds for hundreds of MHz bandwidth photons).

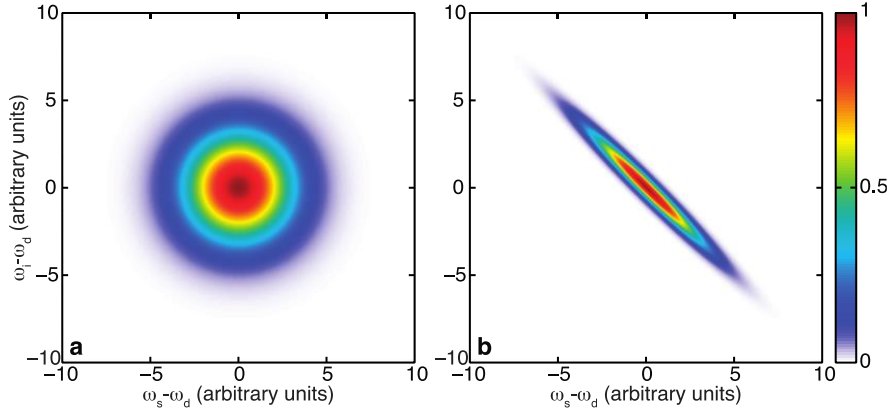

**Figure 3.** Examples of normalised joint spectral density for frequency-uncorrelated (a) or frequency-correlated (b) signal and idler photons. The axes represent the frequency shift with respect to degeneracy ($\omega_d = \omega_p/2$ and $\omega_d = \omega_p$ for SPDC and SFWM, respectively) for signal (x-axis) and idler (y-axis) photons.

*3C. Heralding probability:* Another fundamental parameter is the heralding probability – the probability of measuring a signal photon once the heralding idler counterpart has been detected. This quantity is strictly related to the loss of the system from generation to detection, and for a lossless system, the probability is 100%. It is defined as[76]:

$$\eta_h = \frac{cc}{c_{heralding}\eta_{det}}, \quad (8)$$

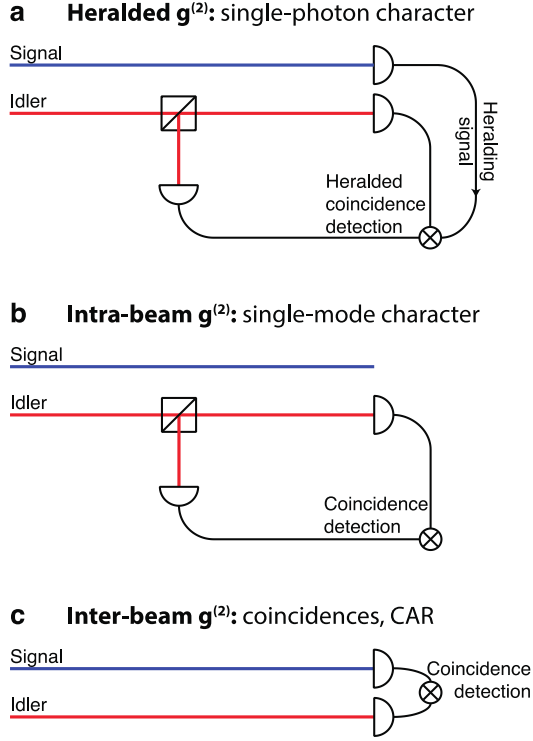

**Figure 4.** Comparison of the experimental setups for measuring the different types of g⁽²⁾ functions reported in this article. The state characterisation allowed by each scheme is also reported.

where *cc* denotes the coincidence counts, $c_{heralding}$ denotes the single counts on the heralding arm (e.g., idler), and $\eta_{det}$ is the quantum efficiency of the detector on the heralded single-photon arm (signal). The heralding probability allows for a comparison of different sources independent of the specific detectors used.

*3D. Coincidence to accidental ratio (CAR):* This parameter characterises how well the source generates photon pairs for both entangled pair and heralded photon sources. It is evaluated by measuring the coincidences between the signal and idler photons as a function of the relative delay ($g^{(2)}_{si}$, often referred to as inter-beam *g⁽²⁾* or intensity cross-correlation (see Fig. 4)). In the ideal case, where signal and idler photons are emitted only in single pairs and without noise or loss, coincidences occur only near zero delay, with no coincidences at all for delays longer than the signal idler coherence time (typically determined by the phase matching conditions for single-pass SPDC and SFWM and by the cavity lifetime for cavity-enhanced processes). The CAR is often defined as:

$$CAR = \frac{g^{(2)}_{si}(0)}{g^{(2)}_{si}(\infty)}; \qquad (9)$$

however, this overestimates the true CAR, and a more formal definition should take into account the finite size of the correlation peak[77]:

$$CAR = \frac{\int_{-\tau_{coh}/2}^{\tau_{coh}/2} g^{(2)}_{si}(t)dt}{\int_{T\infty-\tau_{coh}/2}^{T\infty+\tau_{coh}/2} g^{(2)}_{si}(t)dt}, \qquad (10)$$

which represents the ratio between the sum of all coincidences within the peak and the sum of the coincidences over a temporal window of the same size far from the peak ($T_\infty$ is an arbitrary temporal delay far from the peak). In general, the CAR can be affected by loss, by multiple-pair generation, and by noise in the detector[71,78]. If competing emission processes, such as photoluminescence or Raman scattering, are absent, then the CAR is directly related to the probability of emitting multiple pairs[79] and thus to the suitability of a source for generating heralded single photons.

*3E. Entanglement demonstration:* As mentioned above, different criteria can be exploited to demonstrate entanglement. In general, we can divide these into two classes:

i) those based on the violation of a Heisenberg-like inequality for inferred variances, and
ii) those based on the violation of Bell's inequalities[80].

For integrated sources, the vast majority of publications refer to the second class; thus, we also focus on the second class. We refer the reader to the discussion related to Eq. (C.1) in the Supplementary Information, Sec. C, for further details on the first class.

Bell's inequalities have been proposed as a condition that a quantum theory compatible with the local hidden-variables approach (as suggested by Einstein, Podolsky and Rosen[7]) must verify. The violation of Bell's inequalities is not only a proof of entanglement but also demonstrates the non-local realism of quantum mechanics. For the maximally entangled states that are typically generated in SPDC and SFWM, a violation of Bell's inequality can be exploited as proof of entanglement. We refer the reader to[81] for a detailed description of the relation between entanglement and Bell's inequalities.

A more "operative" expression of Bell's inequalities was proposed in 1969[82]; it relies on measuring the coincidence counts between the two arms (A and B) of a bi-partite entangled state for different detector settings. We consider the expression for polarisation entanglement (which is violated by entangled states)[83]:

$$S \equiv |E(a,b) - E(a,b')| + |E(a',b) + E(a',b')| \leq 2, \qquad (11)$$

where *a*, *a'* and *b*, *b'* represent the settings for the two arms A and B (in this case, corresponding to the angles of polarisers in front of the detectors), respectively, and

$$E(a,b) = \frac{cc(a,b) + cc(a+90°, b+90°) - cc(a, b+90°) - cc(a+90°, b)}{cc(a,b) + cc(a+90°, b+90°) + cc(a, b+90°) + cc(a+90°, b)}, \qquad (12)$$

with *cc(a, b)* being the number of coincidences recorded with the signal and idler polarisers set to *a* and *b*, respectively. The angles that can lead to maximum violation of the CHSH (Clauser, Horne, Shimony, Holt) inequality for polarisation entangled states are *a*=0°, *a'*=45°, *b*=22.5°, and *b'*=67.5°.

A different kind of Bell's inequality that can be exploited for demonstrating energy-time entanglement was described by Franson[84]. This state can be generated by pumping a nonlinear crystal with a CW pump having a coherence time larger than the coherence time of the down-converted photons. Energy-time entanglement is formally equivalent to polarisation entanglement when considering two time bins, where the horizontal and vertical polarisations are replaced by early (E) or late (L) time bins[85] (thus the name time-bin entanglement). This two-mode energy-time entangled state can be generated by sending a pump laser through an unbalanced interferometer and then using the generated double-pulse as the pump for a SPDC or SFWM process [84]. With respect to polarisation entanglement, time-bin entanglement is more suitable for fibre propagation, as it is robust against polarisation fluctuations. Time-bin/Energy-time entanglement can be characterised by means of two unbalanced interferometers, one each for signal and idler photons, with variable phase shifters. A CHSH inequality similar to Eq. (11) also holds in this case, with the angles of the polarisers substituted by the phase of the signal and idler interferometers. For the typical time-bin entangled state ($|EE\rangle + |LL\rangle$), the maximal violation of the CHSH inequality is obtained for $a=\pi/4$, $b=0$, $a'=-\pi/4$, and $b'=\pi/2$[86]. Assuming the same average visibility, $V$, of the coincidence between the output ports of 4 interferometers (s1-i1, s1-i2, s2-i1, s2-i2), the CHSH inequality is violated when $V > 1/\sqrt{2} \approx 0.71$. See Supplementary Information, Sec. C, for a discussion on the relationship between entanglement and non-classical correlations.

*3F. Complex quantum state generation:* While most research on the generation of quantum states addresses standard two-partite bi-dimensional states, such as polarisation entangled (2 dimensions) signal and idler pairs, the ability to generate more complex quantum states will strongly benefit applications in communications and computing. On the one hand, high-dimensional quantum states (so-called "quDits") will increase the amount of information per single photon for quantum communications[55]. On the other hand, cluster states[87], i.e., multipartite entangled states in which each particle is entangled with more than one other particle, have been proposed as a fundamental tool for one-way quantum computing[88]. This novel form of computing relies on complex quantum states and simple measurements rather than a complex set of unitary operations on each qubit, as in the more standard circuit model for quantum computing. While cluster states and quDits have been generated in bulk-optic and free-space approaches (see, e.g.,[89–91] and[3,54–57]), both remain an open challenge in chip form, although recent approaches have come close[92,93], and integrated sources of robust multipartite states based on SFWM have been theoretically predicted[94].

**4) On-chip photon sources**

In this section, we review recent advances in sources of single and entangled photons based on nonlinear processes taking place on an integrated chip. While the development of quantum sources using bulk optics is quite a mature field, a more widespread adoption of quantum technologies will require the miniaturisation of devices towards the chip level. This will reduce cost, footprint, and energy consumption and greatly increase reliability.

We classify these integrated sources according to whether they are based on waveguides or cavities, the latter often being used to enhance the nonlinearity as well as to provide unique characteristics of the generated photons (such as narrow bandwidths). Table 1 compares the performances of state-of-the-art results for single- and paired-photon sources for a range of structures, including microcavities, with a focus on CMOS-compatible integrated chips.

| Structures / Parameters | Silicon | | | Hydex | $Si_3N_4$ |
|---|---|---|---|---|---|
| | Nanowire[95] | Ring[77] | PhC[96] | Ring[97] | Ring |
| Nonlinear coefficient ($W^{-1} m^{-1}$) | 300 | - | 4000 | 0.22[98] | - |
| Q-factor | - | 37,500 | - | 1,375,000 | 2,000,000 |
| Coupled pump average power (mW) | 0.18 | 0.019 | 0.055 | 21 | 3 |
| Collected photon bandwidth (GHz) | 25 | 5.2 | 50 | 0.11 | 0.09 |
| Brightness (pairs $s^{-1}$ $mW^{-2}$ $GHz^{-1}$) | $1.6\times10^5$ | $4.4\times10^8$ | $1.5\times10^6$ | $6.2\times10^3$ | $4.3\times10^8$ |
| CAR | 320 | 602 | 330 | 11 | - |
| $g^{(2)}(0)$ | - | - | 0.09 | 0.14 | - |
| Number of entangled photons | 2[99] | 2[100,101] | 2[102] | 4[92] | 2[103] |

**Table 1. Summary of typical experimental results in various $\chi^{(3)}$ structures.**

*4A. Waveguides*. Most integrated sources of quantum states of light are based on centrosymmetric materials such as silicon, $SiO_2$, silicon nitride ($Si_3N_4$), and silicon oxy-nitride ($SiO_xN_y$), which only have third-order nonlinearities[104]. However, there has also been substantial interest in noncentrosymmetric (or $\chi^{(2)}$) materials such as lithium niobate and III-V semiconductors. While possessing both a $\chi^{(2)}$ and $\chi^{(3)}$, they are referred to as "$\chi^{(2)}$" materials since the second-order response dominates the $\chi^{(3)}$ response. We briefly discuss these platforms first.

While often requiring challenging fabrication processes, III-V semiconductors such as AlGaAs offer many advantages, including exhibiting a $\chi^{(2)}$ response and being a direct bandgap semiconductor that can provide optical gain via electrical pumping. One drawback, however, is that III-Vs lack birefringence; thus, phase matching requires novel techniques such as quasi-phase matching (QPM)[105,106] using, for example, Bragg grating reflection waveguides[107] or quantum well intermixing[108]. Polarisation [109–111], time-bin[112] and energy-time[113] entanglement have been achieved using these methods. Correlated photon pairs have also recently been generated in AlGaAs waveguides by exploiting their $\chi^{(3)}$ nonlinearity[114].

Periodically poled lithium niobate (PPLN) QPM waveguides[115,116] have been used to successfully generate cross-polarised photon pairs[117,118] and polarisation entanglement via direct type II configurations[119] by combining either two type II processes using two different poling periods[120–122] or two type I processes by inserting a half-wave plate[123]. Time-bin entanglement[116,124], quantum state generation and manipulation[125–127], "active" quantum walks through nonlinear waveguide arrays[128–130] and photon triplet generation[131] have also all been demonstrated using this platform. By coating a PPLN waveguide with mirror-like facets, a monolithic OPO-based source of energy-time entangled photons[132] has been demonstrated.

The generation of photon pairs in silicon waveguides was considered theoretically in 2006[133] and demonstrated shortly after[134]. Time-bin[135] and polarisation[136] entangled photons were reported, initially with fibre components (Sagnac loop) and then in fully integrated form[99], exploiting an integrated polarisation rotator to combine two type 0 processes. Initially, pulsed pumps were used to achieve sufficient generation rates, but more recently, continuous wave (CW) pumping has been achieved[137], and this is now common. The co-integration of silicon sources with silica devices such as arrayed waveguide gratings (AWGs) has been proven to be a powerful technique[138].

*4B. Microcavities and Microresonators.* Integrated optical cavities greatly enhance the light-matter interaction by spatially or temporally confining and enhancing the radiation by many orders of magnitude, particularly with resonators having quality factors ($Q = \omega/\Delta\omega$, where $\omega$ is the resonance frequency and $\Delta\omega$ is the resonance width) of $10^6$ or even higher. For both highly nonlinear materials, such as silicon or III-V compounds, and more modestly nonlinear materials, such as $Si_3N_4$ and Hydex, cavities offer extreme enhancements in efficiency that can result in parametric fluorescence with pump power only on the order of microwatts. Furthermore, given their small dimensions, cavities can readily be integrated on a chip with other photonic components.

Microdisc, or microtoroid, resonators confine light in whispering gallery modes and can achieve extremely high quality factors[139]. Silica microtoroids have achieved emission of photon pairs with CAR values > $10^3$ and a spectral brightness surpassing that of PPLN bulk crystal sources[140]. Lithium niobate microtoroids have demonstrated the emission of squeezed light (twin beams) far above the OPO threshold[141], as well as the emission of truly single-mode photon pairs[142].

Photonic crystal (PhC) membrane waveguides, both in silicon and III-V semiconductors, are promising sources of non-classical states of light since they enable extreme light confinement that provides a strong enhancement of optical nonlinearities[143–145]. Line-defect, slow-light, PhC waveguides can reduce the group velocity of light to less than 1/50 of its natural speed while keeping the propagation losses low[146]. Correlated photon-pair generation via slow-light enhanced SFWM has been reported[147–150], as well as heralded photon-pair generation in III-V PhC waveguides[151] and even high-dimensional time-bin entangled photons[102]. These experiments achieved a significant enhancement of pair generation efficiency with a strongly reduced footprint compared with conventional photon-pair sources.

Photonic crystal nanocavities $< \lambda^3$ in size and with very high quality factors provide the ultimate interaction between light and matter[152–154]. Microwatt photon-pair generation via SFWM has been reported in a three PhC coupled cavity designed to yield triple resonances at the pump, signal and idler frequencies in an ultrasmall volume ($\ll \mu m^3$)[155]. While fabrication challenges are significant, these nanocavities are promising, high-efficiency, ultralow power sources of quantum states of light. Recently, single-photon nonlinearities[156,157] were achieved in ultrahigh Q/V nanocavities, with the future promise of integrated single-photon sources operating at room temperature via the photon-blockade effect[158,159].

In ring resonators, perhaps the most widely exploited microcavity in quantum photonics, the SFWM[160,161] efficiency for generating photon pairs using $\chi^{(3)}$ is $\sim \gamma\, Q^3/R^2$ (where $\gamma$ is the waveguide nonlinear parameter, $Q$ is the quality factor and $R$ is the radius[160]). This was experimentally verified for silicon rings with $R$=5-30 μm[162] and highlights the trade-off between volume and Q factor. Ring resonators offer extremely high enhancement, particularly for a triply resonant cavity, which occurs if the total dispersion is low (i.e., within a constant free spectral range, FSR=$v_g/(2\pi R)$, where $v_g$ is the group velocity). Efficient dispersion engineering has been achieved in both silicon and SiN platforms[104]. Initial experiments verified the coincidences between signal and idler photons [102] sent to different single-photon detectors by measuring the inter-beam $g^{(2)}$[137], in which generation rates of $10^5$ Hz with a CAR of 30 were achieved using < 1 dBm CW pump power. A better figure of merit of $10^7$ Hz with a CAR of 50, achieved under the same pumping conditions, was later demonstrated in a 10 μm ring with a $Q$ of $10^4$[162].

Ring resonators are particularly promising sources of time-energy or time-bin entangled states in the telecom band for QKD applications[100,101,163]. Their narrow emission bandwidths, on the order of a few GHz, are compatible with DWDM (dense wavelength division multiplexing) networks, and the required frequency and low power of the pump makes remote pumping possible, with the resulting spectral brightness being comparable to the best second-order nonlinear crystals[100]. In addition, ultrahigh $Q$ resonators yield extremely narrow linewidths, commensurate with quantum memories that typically rely on atomic transitions with linewidths on the order of 100 MHz or less[164]. CROW (coupled-resonator optical waveguide) devices increase the nonlinear parameter by ten times or more[165] and have been shown to be efficient heralded single-photon sources[148], wavelength multiplexed photon-pair sources[166] and time-bin entangled photon[167] sources.

Finally, it has been shown that ring resonators are particularly appealing for heralding single photons in a pure state without the need for external spectral filtering. In fact, when used as a heralded single-photon source, a typical resonator pumped by a field having a spectral width broader than the resonance linewidth can generate heralded single photons with a purity as high as 92%[18,73,160]. Moreover, it has been recently suggested that individual control of the spectral width of the resonances involved in SFWM can lead to fully spectrally unentangled photon pairs; in this case, the purity can theoretically reach 100%[168].

One challenge with SFWM – whether in waveguides or cavities – is that the pump exists in the same spectral region as the generated photon pairs instead of at twice their frequency, as in SPDC. This makes filtering out the pump, which is typically 90-100 dB stronger than the generated signal and idlers, a significant challenge. Very recently, however, this level of rejection was demonstrated on a chip[169] for pair generation[170].

Silicon has, in many ways, been the "workhorse" for quantum applications based on integrated nanophotonics. The use of standard 45 nm CMOS fabrication processes has enabled the integration of ring resonators with electronic components[171] as well as with other optical devices, such as filters, modulators, detectors, and splitters of degenerate photon pairs[172]. However, the moderately high linear (a few dB/cm) and significant nonlinear loss (two-photon absorption – TPA) of silicon have served as limitations, despite the use of novel techniques such as integrating P-I-N junctions to sweep away TPA-generated free carriers to allow higher pump powers to yield higher emission rates of $10^8$ Hz[77].

This has led to the need for developing new nonlinear platforms, including $Si_3N_4$ and Hydex [11], that exhibit both extremely low linear and, perhaps more importantly, low nonlinear optical loss[173,174]). Although Hydex – similar to silicon oxynitride – has a lower nonlinearity than silicon, very high Q ring resonators can be achieved ($> 10^6$), which greatly enhances the SFWM[98,175,176]. The emission of pairs for heralded single-photon sources was demonstrated over a 200 GHz multifrequency comb compatible with the ITU frequency grid for dense wavelength division multiplexed optical networks[97]. This would allow the transmission of quantum states over fibre-optic networks, as well as the use of standard telecom filters to route the different wavelengths and deterministically separate signal and idler photons. The high Q factor yielded photon pairs with narrow linewidths – compatible with quantum memories (~150 MHz). Very recently [140], the emission of entangled photons was also reported, with the multifrequency nature of the emitted signal idler pairs being exploited to enable an on-chip source of four-photon time-bin entangled states[92] (Fig. 5). In moderate refractive index materials such as Hydex, fibre-to-chip coupling can be extremely efficient; this coupling has allowed the use of self-pumping techniques with optical amplifiers to avoid the need for expensive external tuneable lasers, which is important for practical applications[97,177]. Advanced time-bin entanglement circuits have also been reported in ultralow-loss silicon nitride photonic chips[178]. Recently, Hydex micro-ring resonators achieved type II SFWM on a chip by exploiting subtle birefringent effects, thus paving the way for the direct generation of polarisation entanglement on a chip in a single process[179]. Silicon nitride ($Si_3N_4$) ring resonators are also very interesting candidates for generators of quantum optical states[180], including entangled photon pairs[103], twin beams[181,182], and random numbers[183].

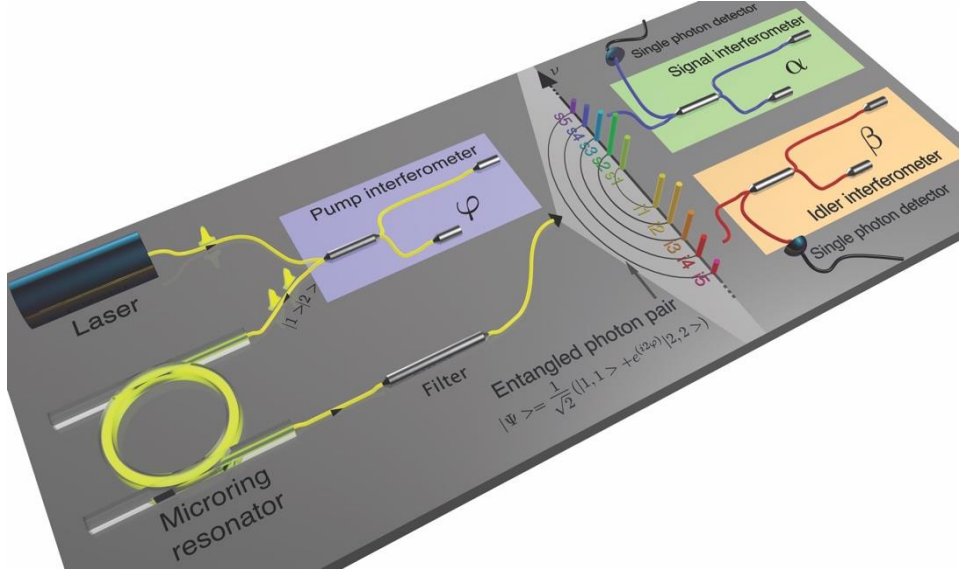

**Figure 5.** Quantum frequency comb generation and detection setup based on time-bin entanglement in a ring resonator[92]. A pulsed laser (16.8 MHz repetition rate passively mode-locked fibre laser with a bandwidth of 0.1 nm, spectrally centred at 1556.2 nm) is passed through an unbalanced Michelson interferometer (consisting of a 50/50 beam splitter, Faraday mirrors, and a phase shifter), generating two pulses with a phase difference φ in two respective time slots (time bins |1⟩ and |2⟩). The pulses are fed into the micro-ring resonator (see arrows for the propagation direction), exciting one micro-ring resonance. The nonlinear spontaneous four-wave mixing process generates signal-idler photon pairs on several ring resonances symmetric to the excited resonance (optical frequency comb, indicated in multicolour), either within the first or the second time slot (the generation in both time bins is made highly improbable by the chosen low excitation power). The superposition of the state generated in the first and the second time slot results in an entangled state output |ψ⟩, which takes place simultaneously on several resonances and leads to a frequency comb of time-bin entangled photon pairs. For analysis purposes (entanglement verification or quantum state tomography), each photon of the spectrally filtered photon pair (distributed on two resonances symmetric to the excitation frequency, e.g., the resonance pair i4-s4 used here) is individually passed through an interferometer, with the temporal imbalance equal to the time slot separation, and then detected using a single-photon detector (note that the phases α and β of the second and third interferometers can be individually controlled).

## 5) Deterministic sources

Deterministic photon sources are desired for many applications, such as quantum computing and communications, since the interaction probability between multiple single photons from independent random sources is far too low to be practical. While non-classical emitters such as quantum dots[184–186] or nitrogen vacancies in diamonds[187] can produce single photons deterministically and are promising sources, they are not without their challenges. Photon collection losses can degrade their deterministic nature, and even though photons created from the same emitter show very high indistinguishability[184,185], achieving enough uniformity with nanoscale accuracy[186,187] to generate indistinguishable photons from multiple emitters is difficult, often requiring narrowband filtering[186].

Photon generation via nonlinear optics also has its challenges, as it is intrinsically random, being governed by statistical distributions (e.g., Poissonian and thermal) that limit the single-photon generation probability to less than 25%[188]. However, "heralding" can increase the probability of single-photon generation without sacrificing the source quality through the use of, for example, active multiplexing techniques[150,189–195]. More importantly, photons from separate nonlinear sources have been shown to be highly indistinguishable[195].

Photon multiplexing can be achieved in space[150,189,190] or time[191–195]. Fig. 6 shows two multiplexing schemes that can actively combine heralded photons from $N$ different modes (in this case, $N$=4). In spatial multiplexing, as shown in Fig. 6a, correlated photon pairs are randomly generated in some of the waveguides via SFWM. One and only one heralded photon is routed to the output at a time according to predefined logic in a field-programmable gate array (FPGA); thus, the single-photon output probability is enhanced[150]. This scheme, however, requires many devices for each photon source and thus is difficult to scale up. Temporal multiplexing, as illustrated in Fig. 6b, is much more efficient because only one photon source is required and the photons to be multiplexed are generated from different temporal modes. When photons from 4 modes are multiplexed, the enhancement of the single-photon output probability is 100%, and the Hong-Ou-Mandel (HOM[196]) interference with the multiplexed photons exhibits 91% visibility[195]. So far, however, the single-photon generation efficiency after multiplexing has been very low. This is mainly because the starting point for multiplexing – the source before multiplexing – has to operate in the low efficiency regime to avoid multiphoton noise. If photon-number-resolving detectors[197] can be exploited, one can start at the theoretical limit of 25% single-photon generation probability and use scalable temporal multiplexing schemes to achieve nearly deterministic single-photon sources. Of course, the overall loss, including, in particular, the loss due to the switches[195], is a critical factor since this can significantly degrade the overall fidelity of a single-photon source.

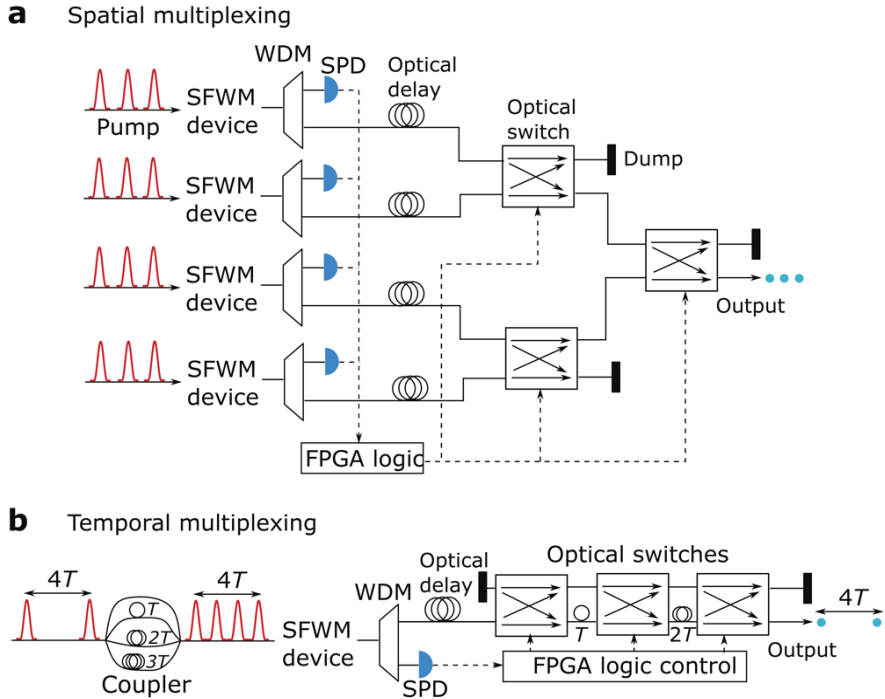

**Figure 6.** The multiplexing schemes. WDM: wavelength division multiplexing, SPD: single-photon detector, FPGA: field programmable gate array.

**6) Conclusions**
We review the current state-of-the-art in photonic integrated circuits designed to generate complex photonic quantum states, focusing on devices based on nonlinear optics that are compatible with quantum memories, with fibre optic communications, as well as with silicon integrated circuit semiconductor technology (CMOS). These new developments play a key role in realising compact, low-cost, and practical sources of complex quantum optical states on a chip, which will ultimately enable quantum technologies to have a significant impact on our society.

**Acknowledgements**
This work was supported by the Natural Sciences and Engineering Research Council of Canada (NSERC) through the Steacie Memorial Fellowship. D.J.M. was supported through the Australian Research Council Discovery Projects programme (DP150104327). L.C. acknowledges the support of the People Programme (Marie Curie Actions) of the European Union's FP7 Programme under REA Grant Agreements No. 627478 (THREEPLE). B.J.E. and C.X. acknowledge the Australian Research Council (ARC) Centre of Excellence (CUDOS, CE110001018), Laureate Fellowship (FL120100029), and the Discovery Early Career Researcher Award (DE120100226) programmes.

# Supplementary Information

**A: Definition of quantum superposition and quantum entanglement**
*Quantum superposition:* If we consider, for example, a quantum state that can be described by a two-dimensional space (such as a photon polarisation or an electron spin), we find that the quantum state can be described not only as being in state A (e.g., horizontal polarisation or spin up) *or* B (e.g., vertical polarisation or spin down) but also in a superposition of the two:

$$|\psi\rangle = \alpha|A\rangle + \beta|B\rangle, \quad (A.1)$$

where $\alpha$ and $\beta$ are two complex parameters related by the normalisation condition $|\alpha|^2 + |\beta|^2 = 1$. Note that this is conceptually different from the scenario where a system is either in state *A* with probability $|\alpha|^2$ *or* in state *B* with probability $|\beta|^2$ (a state known as the mixed state and often represented as $\{|A\rangle; |B\rangle\}$).

*Quantum entanglement:* Superposition also applies to the quantum state of two separate systems (such as two photons and two electrons). These two systems are said to be entangled if their state cannot be described separately, e.g.:

$$|\psi\rangle_{ent} = \alpha|A\rangle_1|A\rangle_2 + \beta|B\rangle_1|B\rangle_2, \quad (A.2)$$

which represents a superposition in which the two systems are in both state A *and* state B. In contrast, a separable state can always be described as the product of the independent states of systems 1 and 2, e.g.:

$$|\psi\rangle_{sep} = |\psi\rangle_1 \otimes |\psi\rangle_2 = (\alpha|A\rangle_1 + \beta|B\rangle_1) \otimes (\gamma|A\rangle_2 + \delta|B\rangle_2) . \quad (A.3)$$

**B: Multimode emission and mixed state in heralded single-photon sources**
In SPDC and SFWM, the strong correlations between signal and idler photons determined by momentum and energy conservation can lead to multimode emission. For example, we can consider the frequency correlations between signal and idler photons, where the state can be expressed as:

$$|\psi\rangle_{SPDC/SFMW} = |\omega_1\rangle_s|\omega_{-1}\rangle_i + |\omega_2\rangle_s|\omega_{-2}\rangle_i + |\omega_3\rangle_s|\omega_{-3}\rangle_i + \cdots, \quad (B.1)$$

where frequency pairs $\omega_n$ and $\omega_{-n}$ sum up to the pump frequency for SPDC (or twice the frequency pump for SFWM): $\omega_n + \omega_{-n} = \omega_{pump}$ (or $2\omega_{pump}$). For simplicity, we consider the case of discrete frequencies, as would be the case for SPDC or SFWM in a cavity; however, similar results would also be obtained when taking into account the continuous character of the frequency distribution. Measuring the heralding photon (say the idler) *without resolving its frequency* projects the signal photon into a mixed state $|\psi\rangle_{signal} = \{|\omega_1\rangle; |\omega_2\rangle; |\omega_3\rangle; ...\}$. A possible solution is to filter only a single frequency mode so that the state is, e.g., $|\psi\rangle_{SPDC/SFMW} = |\omega_2\rangle_s|\omega_{-2}\rangle_i$. In this case, measuring the heralding idler photon will project the single photon into the pure state $|\omega_2\rangle_s$.

**C: Relation between non-classical correlations and entanglement**

Entanglement and non-classical correlations, which are commonly interchanged, are in fact quite different. While the presence of non-classical correlations is not enough to demonstrate entanglement, the reverse is true – entanglement guarantees the presence of non-classical correlations. To better clarify the difference, we consider a practical example such as the so-called "twin beams". In this case, as the name suggests, two beams are said to be twins if they display exactly the same intensity at the single-photon level. For example, a laser field impinging on a lossless perfectly balanced 50/50 beam splitter will not generate twin beams; indeed, the intensity at the output ports of the beam splitter is the same only on average. The two beams will exhibit non-correlated intensity fluctuations determined by the quantum nature of light (shot-noise). The amplitude of these fluctuations scales as the inverse square root of the average intensity. On the other hand, in both the SPDC and SFWM processes, in the ideal scenario, the signal and idler beams generated will exhibit the exact same photon statistics. This is intrinsic to the generation process, as one signal photon can be generated if and only if an idler photon is also generated.

Indeed, one means for proving the entanglement of a bipartite system is based on the Peres-Horodecki criterion[198,199]. It defines a necessary condition for separability – its violation is a sufficient (but not necessary) condition for entanglement[3]. However, the violation of this criterion is a sufficient and necessary condition only for 2×2 and 2×3 systems (bipartite two-mode and bipartite tri-mode systems, respectively). We note that the Peres-Horodecki criterion is more sensitive than Bell's inequalities in the sense that there exist states that are entangled according to the Peres-Horodecki criterion that do not violate any of Bell's inequalities[200,201].

A version of the Peres-Horodecki criterion for continuous variables was proposed in 2000 by Duan[202] and Simon[203] and shows that for an entangled state the inferred variances of two non-commuting variables, denoted by the operators $\hat{p}$ and $\hat{q}$ (e.g., energy/time, intensity/phase, and position/momentum), violate an inequality of the form:

$$V(\hat{p}_1 - \hat{p}_2) + V(\hat{q}_1 + \hat{q}_2) \geq 2\sqrt{2}, \qquad (C.1)$$

where $V(\hat{x}) = \langle\hat{x}^2\rangle - \langle\hat{x}\rangle^2$. Equation (C.1) has been used, for example, to demonstrate the entanglement in the case of twin beams[204,205].

Therefore, in the case of twin beams, entanglement can be demonstrated by showing non-classical correlations between the beam intensities *and* phases[204,205]. However, for well-known systems, such as SPDC and SFWM, for which we know that the origin of the non-classical correlations is indeed entanglement, the existence of these correlations is often considered as an indication of the presence of entanglement. It is also important to stress that non-classical correlations are also used as the basis for quantum metrology, where, for many applications, entanglement is not required since a reduction in the noise of one of the two variables is necessary to achieve higher sensitivity in, for example, high-sensitivity quantum spectroscopy and imaging[206-209].